\documentclass[12pt]{article}
\usepackage{amsfonts}
\usepackage{times}

\def\co{\Delta}  
\def\Dco{\Delta}

\def\be{\begin{equation}}
\def\ee{\end{equation}}
\def\bea{\begin{eqnarray}}
\def\eea{\end{eqnarray}}

\def\co{\Delta}

\def\>#1{{\bf #1}}

\def\1{\'{\i}}                           
\def\R{\rm I\kern-.2em R} 
\def\C{\rm I\kern-.5em C} 
\def\back{\!\!\!\!\!\!}

\def\tfrac#1#2{ {\scriptstyle { \frac {#1}{#2}}}}         
\def\pois#1#2{\left\{ {#1},{#2} \right\}}         
\def\poiss#1#2{\left\{ {#1}\otimes_{\ _{\!\!\!\!\!\!\!,}}{#2} \right\}}         
\def\conm#1#2{\left [ {#1},{#2} \right ]}         
\def\poisq#1#2{\left\{ {#1},{#2} \right]}

\begin{document}
\thispagestyle{empty}
\hfill\today 
\vspace{2.5cm}

 \

\begin{center}

{\Large{\bf{Classical Dynamical Systems from
$q$-algebras:}}} 
\smallskip

{\Large{\bf{``cluster" variables and explicit solutions}}} 
\smallskip

\vspace{1.4cm} 

{\large {\sc \'Angel Ballesteros}}\vspace{.2cm}\\
{\it Departamento de F\1sica, Universidad de Burgos\\
Pza. Misael Ba\~nuelos s.n., E-09001-Burgos, Spain} 
\vspace{.3cm}

{\large {\sc Orlando Ragnisco}}\vspace{.2cm}\\
{\it Dipartimento di Fisica,  Universit\`a di Roma TRE\\
Via della Vasca Navale 84, I-00146-Roma, Italy} 
\vspace{.3cm}

\end{center} 
  
\bigskip

\begin{abstract} 
A general procedure to get the explicit solution of the equations of motion for
$N$-body classical Hamiltonian systems equipped with coalgebra symmetry is introduced
by defining a set of appropriate collective variables which are based on the iterations
of the coproduct map on the generators of the algebra. In this way several examples of
$N$-body dynamical systems obtained from
$q$-Poisson algebras are explicitly solved: the $q$-deformed version of the $sl(2)$
Calogero-Gaudin system ($q$CG), a
$q$-Poincar\'e Gaudin system and a
system of Ruijsenaars type  arising from the same (non co-boundary)
$q-$deformation of the (1+1) Poincar\'e algebra. While the complete integrability of
all these systems was already well known, being in fact encoded in their construction,
no explicit solution was available till now. In particular, it turns out that there exists an open subset of the whole phase space where the
orbits of the 
$q$CG system are periodic with the same period. Also, a unified 
interpretation of all these systems as different Poisson-Lie dynamics on the
same three dimensional solvable Lie group is given.
\end{abstract}

\vfill
\newpage

\section{Introduction}

The coalgebra approach \cite{Ragn}-\cite{Rag}, and more recently its
comodule algebra generalization \cite{BMR}, have been proven to be quite effective
methods to construct both classical and quantum  hamiltonian systems with a large
family of commuting integrals of the motion. In some important cases (e.g.
for rank 1 Lie Algebras and their $q$-analogues)  the coalgebra method leads to the
proof of the complete integrability and even of the ``super-integrability" 
\cite{BalHerMont} of such kind of systems. Moreover, in the quantum-mechanical case
both non deformed and
$q$-deformed systems  can be handled and explicitly solved on the same footing
\cite{MR1,MR2}.

Surprisingly enough, such an explicit solution looked harder to get at the
classical (Poisson) level, inasmuch as some traditional ingredients of classical
integrability (such as the Lax representation) were missing for all these systems. This
issue has been recently addressed in \cite{GroupTheoryConf}, where some steps towards
the solution of the $q$CG system were accomplished.

In this paper we go ahead along the
lines  introduced in \cite{GroupTheoryConf}  and we present a general approach to the explicit solution of
$N$-body classical Hamiltonian systems with coalgebra symmetry.
This is achieved in the next Section by using the coalgebra structure in order to define
a set of collective variables (the so-called ``cluster" variables) whose equations of
the motion can be firstly separated by making use of the integrals for the system, and
finally solved. Essentially, the coalgebra
symmetry implies a kind of self-similarity of the dynamical equations for these cluster
variables, and the
$N=2$ case turns out to be the essential cornerstone in order to solve the full
$N$-body dynamics. 

 This general procedure is applied to several coalgebra systems
in Section 3. There we exhibit the complete solution of the
classical equations of motion for the 
$q$CG system, which has the $N$-body Casimir function as the Hamiltonian, unveiling the
deep connection between non deformed and $q$-deformed systems. Afterwards, two
more systems defined on a non-coboundary $q$-deformation of the (1+1) Poisson Poincar\'e
algebra are explicitly solved: the first one is a $q$-Poincar\'e analogue of the CG
system and the second one is the Ruijsennars-Schneider like system introduced in
\cite{Rag}. 

Finally, Section 4 presents  a unified  interpretation of all the previous
systems as different cases of dual Poisson-Lie dynamics on $N$ copies of the three
dimensional solvable Lie group $G_z$ generated by two translations and one dilation.
From this point of view the coproduct map (which is the formally same for all the
systems here considered) is just the Lie group multiplication of $G_z$, and the
$m$-th cluster variables are essentially the entries of the matrix representation of
the product of $m$ different $G_z$ group elements. Therefore, the $q$CG system and the
two $q$-Poincar\'e systems here solved are just two particular cases among the set of
all dynamical systems defined by the Poisson-Lie structures on $G_z$. All such systems
are, by construction, compatible with the $G_z$ group multiplication but exhibit quite
different dynamical features, as it will be shown through the examples here introduced.

\section{Coalgebras and cluster dynamics}

We recall from \cite{Rag} that the following general result can be
proven: 

\noindent
{{Let $(A,\Delta^{(2)})$ be a (Poisson) coalgebra with generators $X_k$
($k=1,\dots,l$) and with Casimirs
$C_i$ ($i=1,\dots,r$). Then, the
$m^{th}$-coproducts of the Casimirs (Poisson)-commute among themselves and with the
$N^{th}$-coproducts of the generators of $A$}}:
\bea
\poisq{\Delta^{(m)}(C_i)}{\Delta^{(n)}(C_j)} = 0 \quad m,n =1, \dots, N \quad
\forall\  i,j ;\\
\poisq{\Delta^{(m)}(C_i)}{\Delta^{(N)}(X_k)} =0 \quad m=1,\dots N \quad
\forall\  
k,\quad \forall\  i.
\nonumber
\eea
Here the symbol $\poisq{\cdot}{\cdot}$ simultaneously denotes both the commutator
(for Quantum systems where the where the algebra of observables
$A$ is non-commutative) and its Poisson analogue for Classical Mechanical systems.
Nothe that with this notation the
$\Delta^{(1)}$ coproduct is just the identity map. Hence, for rank $r=1$  coalgebras
(and their Poisson analogues), we can consider as a Hamiltonian any (smooth) function 
$H$  of the three generators of the algebra:
\be
H=H(X_1,X_2,X_3),
\ee
and the complete integrability is thus established for the
Hamiltonian $H^{(N)}$ defined as the
$N^{th}$-coproduct of $H$:
\be
H^{(N)} = H(\Delta^{(N)} (X_1),\Delta^{(N)} (X_2),\Delta^{(N)} (X_3)). 
\ee
Note that the Casimir $C$ is just a particular function of the generators, and we could
take $H\equiv C$ as a particular type of system (the so-called Gaudin-type
Hamiltonians).

As it was extensively
explained in \cite{Rag}, the proof of the abovementioned statement relies on the fact
that the coproduct $\Delta^{(2)}$ is a coassociative homomorphism: this means that
there exist two equivalent ways to define the $3$-rd order coproduct map, namely
\be  \co^{(3)}:=(id\otimes\co^{(2)})\circ\co^{(2)}=(\co^{(2)}\otimes id)\circ\co^{(2)}
\label{fj}
\ee
and, in general, $(N-1)$ equivalent definitions of the $N$-th order coproduct are
possible:
\be
\co^{(N)}:=(\co^{(m)}\otimes \co^{(N-m)})\circ\co^{(2)},
\qquad  m=1,\dots,N-1.
\label{buena}
\ee
This property holds
for the Poisson case as well as in the quantum context, both in the undeformed and
in the $q$-deformed case, and it underlies the superintegrability properties of
the dynamical systems with coalgebra symmetry \cite{BalHerMont,BRsuperint}.

In this paper we address the problem of the explicit solution of the classical dynamics
of such $N$-body Hamiltonian systems. For that
purpose, the choice of an appropriate set of dynamical variables will be essential. In
this respect, we can rewrite the
expression (\ref{buena}) by using Sweedler's
notation \cite{Sweed} in which the coproduct
$\co^{(2)}$ of an arbitrary element $Y$ of the algebra $A$ (with arbitrary rank $r$) is
given as the linear combination
\be
\co^{(2)}(Y)=\sum_{\alpha}{Y_{1\,\alpha}\otimes
Y_{2\,\alpha} },
\label{Swee}
\ee
where $Y_{1\,\alpha}$ and $Y_{2\,\alpha}$ will be certain functions that live on two
diffrent copies of the algebra $A$:
\be
Y_{1\,\alpha}=X_{1\,\alpha}(X_1,\dots,X_l),\qquad 
Y_{2\,\alpha}=X_{2\,\alpha}(X_1,\dots,X_l).
\ee
From (\ref{buena}) we immediately get that, in this notation, the $N$-th
coproduct of $Y$ reads
\be
\co^{(N)}(Y):=\sum_{\alpha}{ \co^{(m)}(Y_{1\,\alpha})\otimes
\co^{(N-m)}(Y_{2\,\alpha}) }, \qquad  m=1,\dots,N-1,
\label{buena2}
\ee
and since any $\co^{(p)}$ map is an algebra homomorphism we can write 
\be
\co^{(N)}(Y):=\sum_{\alpha}{ Y_{1\,\alpha}(\Delta^{(m)}
(X_1),\dots,\Delta^{(m)} (X_l))\otimes
Y_{2\,\alpha}(\Delta^{(N-m)}
(X_1),\dots,\Delta^{(N-m)} (X_l)) }. 
\label{buena2}
\ee
where $m=1,\dots,N-1$.

The expression (\ref{buena2}) leads in a natural way to the definition of the
following set of
$2\,l\,(N-1)$ collective ``cluster" variables
\be
X_k^{(m)}:=\Delta^{(m)}(X_k) \qquad\qquad
X_k^{(N-m)}:=\Delta^{(N-m)}(X_k) 
\label{cluster}
\ee
where $k=1,\dots,l \quad \mbox{and}\quad m=1,\dots N-1$. With them (\ref{buena2}) is
rewritten as
\be
\co^{(N)}(Y):=\sum_{\alpha}{ Y_{1\,\alpha}(X_1^{(m)},\dots,X_l^{(m)})\otimes
Y_{2\,\alpha}(X_1^{(N-m)},\dots,X_l^{(N-m)}) }
\label{buena3}
\ee
for any fixed value of $m$.
We stress that the variables
$X_k^{(N-m)}$ live on the $(N-m)$ tensor copies of $A$ which are located starting from
the right of the full chain of $N$ copies of $A$:
$$\overbrace{A\otimes A\otimes\dots\otimes  A}^{m}\otimes 
\overbrace{A\otimes\dots \otimes A\otimes  A}^{N-m}$$
 Therefore, 
\be
\pois{X_k^{(m)}}{X_p^{(N-m)}}=0,\qquad \forall\ k,p.
\ee
Moreover, since the coproduct is always a Poisson algebra homomorphism, the set of
variables $\{X_1^{(m)},\dots,X_l^{(m)}\}$ reproduces again the Poisson algebra $A$, and
the same is true for the complementary set $\{X_1^{(N-m)},\dots,X_l^{(N-m)}\}$. In this
notation the $m$-th coproducts of the Casimirs, which are constants of the motion for
$H^{(N)}$, are written as
\be
C^{(m)}=\Delta^{(m)}(C)=C(X_1^{(m)},\dots,X_l^{(m)}).
\ee
By construction, there also exists a complementary set of integrals in involution
defined by
\be
C^{(N-m)}=\Delta^{(N-m)}(C)=C(X_1^{(N-m)},\dots,X_l^{(N-m)}).
\ee
These additional integrals explicitly show  the intrinsic superintegrability properties
of coalgebra symmetric systems \cite{BalHerMont,BRsuperint}.

Now, the Hamiltonian $H^{(N)}$ is just a particular case of (\ref{buena2}),
and we can write it in $(N-1)$ equivalent ways labeled by $m$:
\be
H^{(N)}=H\left(X_1^{(m)},\dots,X_l^{(m)};X_1^{(N-m)},\dots,X_l^{(N-m)}\right)
\qquad m=1,\dots,N-1.
\ee
From this perspective it is clear that the dynamics of the system will be
explicitly known if we are able to solve the evolution equations for the cluster
variables $X_k^{(m)}$, which are
$(N-1)$ sets of $l$ nonlinear coupled first order ODE's
\be 
\dot{X}_k^{(m)}=\pois{X_k^{(m)}}{H^{(N)}}=
f_k (X_1^{(m)},\dots,X_l^{(m)};X_1^{(N-m)},\dots,X_l^{(N-m)})
 \label{systemm}
\ee
where, in turn, the complementary cluster variables $X_k^{(N-m)}$ fulfill the equations
\be
\dot{X}_k^{(N-m)}=\pois{X_k^{(N-m)}}{H^{(N)}}=
g_k (X_1^{(m)},\dots,X_l^{(m)};X_1^{(N-m)},\dots,X_l^{(N-m)}).
\label{system2}
\ee
As a final step, we will also have to solve the $l$ coupled equations that provide the
dynamics of the $N$-th coproducts of the generators:
\be
\dot{X}_k^{(N)}=\pois{X_k^{(N)}}{H^{(N)}}=
h_k (X_1^{(N)},\dots,X_l^{(N)}).
\label{systemn}
\ee
Note trhat the dynamics of the latter $N$-body collective cluster variables 
turns out to be different from any other lower $m$-th dimensional $m<N$
set of cluster variables. 

Despite the apparent complexity of the problem, we stress that all the
$(N-1)$ sets of equations
(\ref{systemm}) are formally identical for any $m=1,\dots,N-1$, since the functions
$f_k$ (and $g_k$) do not depend on $m$ (this follows from the fact that the
coproduct is a homomorphism). As a consequence, the dynamics of
the collective cluster variables is the same whatever the number $m$ of degrees
of freedom of the cluster is. This ``self-similar" dynamical behaviour turns out to be
an absolutely general feature of Poisson coalgebras, and it holds for both non-deformed
and deformed coalgebra symmetric systems. In the latter case the
cluster structure of the dynamics is preserved by the $q$-deformation, but the
(always long-range) interactions coming from the
$q$-deformation implies a more involved dynamics.

As we shall see in the following sections, the
strategy followed in order to solve the system (\ref{systemm}) will be to eliminate the
complementary variables
$X_k^{(N-m)}$ by making use of the constants of the motion given by the coalgebra
symmetry (therefore, without working out the explicit solution for (\ref{system2})). In
particular, a distinguished set of dynamical systems appears when the Hamiltonian
function $H^{(N)}$ is just the $N$-th Casimir
$C^{(N)}$ (the so-called Gaudin Hamiltonians). In this particular case, since the Poisson brackets
(\ref{systemn}) vanish,
the
$N$-th coproducts of the generators
$X_k^{(N)}$ give a set of $l$ constants of the motion that we shall call $\delta_k$.
Moreover,  the expressions of the
$N$-th coproducts
$X_k^{(N)}$ obtained by using (\ref{buena2}) give rise to
$l$ different algebraic equations that, for a given $m$, involve both the $X_k^{(m)}$
and the
$X_k^{(N-m)}$ variables. When the latter are eliminated from them by using such
constants, the system (\ref{systemm}) is transformed into
\be
\dot{X}_k^{(m)}=\pois{X_k^{(m)}}{H^{(N)}}=
\tilde{f}_k (X_1^{(m)},\dots,X_l^{(m)};\delta_1,\dots,\delta_l)
\ee
which represents $m$ copies of the same
one-body dynamics in an ``external'' field (para\-metrized by $\delta _1,\dots,\delta
_l$) and thus can be simultaneously solved for any value of
$m$. In case we deal with more general Hamiltonians different from the $N$-th Casimir,
the lower dimensional ``cluster Casimirs"
$C^{(m)}$ and $C^{(N-m)}$ will provide alternative integrals that can be used in order
to eliminate the $X_k^{(N-m)}$ variables from the equations (\ref{systemm}).

Finally, we recall that in all the previous constructions of 
coalgebra systems (see for instance \cite{Balrag}), the Hamiltonians have been written
in terms of canonical coordinates coming from specific symplectic realizations of the
algebra
$A$:
\be
X_k=X_k(q_1,\dots,q_n;p_1,\dots,p_n;c_1,\dots,c_r)\qquad k=1,\dots,l
\label{can}
\ee
where $c_i$ are the values of the Casimirs of $A$ that define the symplectic manifold
on which the functions (\ref{can}) close the algebra $A$. The number
$n$ of pairs of canonical variables needed in order to get such a realization depends
both on the number of generators $l$ of
$A$ and of its rank $r$. For $r=1$, it turns out that $n=1$ and then the cluster
variable $X_k^{(m)}$ would be a function of just $m$ canonical pairs.  However, for
higher rank algebras
$X_k^{(m)}$ will depend on a number
$n\cdot m$ of canonical coordinates and momenta, and the equations (\ref{systemm}) will
give us the ``global" dynamics of the cluster variables encompassing
simultaneously $n\cdot m$ degrees of freedom, whose individual time evolution can only
be obtained by inverting (if possible) the algebraic equations
(\ref{can}).

\section{Explicit solutions of coalgebra systems}

\subsection{The non-deformed CG system}

As a first ``toy model" with $sl(2)$ coalgebra symmetry, we will solve the dynamics of
the Calogero-Gaudin system \cite{Gaudin1}-\cite{Kar} by using the approach of the
previous Section. 

Let us consider the $sl(2)$  Poisson coalgebra
\be
\pois{X_3}{X_\pm} = \pm\,2\, X_\pm  \qquad \pois{X_+}{X_-} = X_3
\ee 
equipped with the primitive coproduct 
\be
\Delta (X_i) = X_i \otimes id + id \otimes X_i \qquad\quad i=3,\pm
\label{prim}
\ee
and with Casimir function:
\be
C = \frac{1}{4}X_3^2 + X_+ \, X_-.
\ee
We consider the $N$-th coproducts of the generators
\bea
&&\!\!\!\!\!\!\!\!\!\!\!\!\!\!X_i^{(N)}:=\Delta^{(N)}(X_i)=X_i\otimes\overbrace{
id\otimes id\otimes
\dots\otimes id}^{(N-1)} +\cr
&& \qquad + id\otimes X_i\otimes \overbrace{id\otimes\dots\otimes
 id}^{(N-2)} +
\dots + \overbrace{id\otimes
id\otimes\dots\otimes id}^{(N-1)}\otimes X_i.
\label{fo}
\eea
and the non-deformed CG system is defined by the Hamiltonian
\be
H^{(N)}:=\Delta^{(N)}(C)=C^{(N)}= \frac{1}{4}(X_3^{(N)})^2 + X_+^{(N)}\, X_-^{(N)}.
\ee
The integrals of the motion for $H^{(N)}$ are, by construction \cite{Balrag}, the $N$-th
coproducts of the three generators
$ 
X_i^{(N)}
$ 
and the $m$-th coproducts of the Casimir 
\be
\qquad C^{(m)}=\frac{1}{4}(X_3^{(m)})^2 + X_+^{(m)}\, X_-^{(m)} \qquad\quad m=1,\dots
(N-1) 
\label{casm}
\ee
together with the complementary integrals
\be
\qquad C^{(N-m)}=\frac{1}{4}(X_3^{(N-m)})^2 + X_+^{(N-m)}\, X_-^{(N-m)} \qquad\quad
m=1,\dots (N-1) 
\label{casNm}
\ee
leading to the well-known superintegrability of the CG system.

An essential property of the primitive coproduct (\ref{fo}) is that, by using
(\ref{buena2}), it can be expressed in $(N-1)$ equivalent forms  in terms of the cluster
variables (\ref{cluster}):
\be
X_i^{(N)}=X_i^{(m)}+X_i^{(N-m)}, \qquad\quad m=1,\dots (N-1).
\ee
The time evolution of such cluster variables is given by (\ref{systemm}) and reads: 
\bea
&& {\dot X}_3^{(m)}=2\,(X_+^{(m)}\,X_-^{(N-m)}- X_-^{(m)}\,X_+^{(N-m)})\cr
&& {\dot X}_+^{(m)}=X_3^{(m)}\,X_+^{(N-m)}- X_+^{(m)}\,X_3^{(N-m)} \label{CG}\\
&& {\dot X}_-^{(m)}=-X_3^{(m)}\,X_-^{(N-m)}+ X_-^{(m)}\,X_3^{(N-m)}
\nonumber
\eea
Now by using the integrals of the motion given by the $N$-th coproducts of the
generators
\be
X_i^{(N)}=X_i^{(m)}+X_i^{(N-m)}\equiv\delta_i,\qquad i=3,\pm
\ee
the variables $X_i^{(N-m)}$ can be eliminated and equations (\ref{CG}) are transformed
into the linear system:
\bea
&& {\dot X}_3^{(m)}=2\,(X_+^{(m)}\,\delta_- -  X_-^{(m)}\,\delta_+)\cr
&& {\dot X}_+^{(m)}=X_3^{(m)}\,\delta_+  - X_+^{(m)}\,\delta_3 \label{CG2}\\
&& {\dot X}_-^{(m)}=-X_3^{(m)}\,\delta_- + X_-^{(m)}\,\delta_3
\nonumber
\eea
The motion for this system can be either hyperbolic or periodic, since the eigenvalues
of the previous system are related to the value of the $N$-th coproduct of the
Casimir in the form
\be
\omega=\pm\,2\,\sqrt{\frac{1}{4}\delta_3^2 +
\delta_+\,\delta_-}=\pm\,2\,\sqrt{C^{(N)}}.
\ee
Note that this result is valid for any ``size" $m$ of the cluster
variables. Finally, we point out that equations (\ref{CG2}) (and therefore the
integration constants) are not independent, since they are constrained by the
$m$-cluster Casimir (\ref{casm}).

\subsection{Dynamics of the $q$-Calogero-Gaudin system}

The ``standard"  $q$-deformation of $sl(2)$ is given by the Poisson brackets
\be
\pois{X_3}{X_\pm} = \pm \,2\,X_\pm  \qquad \pois{X_+}{X_-} =
 \frac{\sinh(zX_3)}{z}
\ee
equipped with the $q$-deformed coproduct $(q=e^z)$
\bea
&& \Delta (X_3) = X_3 \otimes id + id \otimes X_3 \cr
&& \Delta (X_\pm) = e^{-\frac{z}{2}  X_3} \otimes X_\pm + X_\pm \otimes
e^{\frac{z}{2} X_3}.
\label{codef}
\eea
The $q$-deformed Casimir function reads:
\be
C_z =\frac{1}{4}\left\{ \frac{\sinh(\frac{z}{2} X_3)}{z/2}\right\}^2 + X_+ X_-
\ee 

The $N$-th coproduct $\Delta^{(N)}(X_3)$ is just of the type (\ref{fo}). For the
$X_\pm$ generators we have
\bea
&&\Delta^{(N)}(X_\pm)=X_\pm\otimes \overbrace{e^{\tfrac{z}{2} X_3}\otimes
e^{\tfrac{z}{2} X_3}\otimes\dots\otimes  e^{\tfrac{z}{2}
X_3}}^{(N-1)}\cr
&&\qquad\qquad\quad\quad + e^{-\tfrac{z}{2} X_3}\otimes X_\pm\otimes
\overbrace{e^{\tfrac{z}{2} X_3}\otimes\dots\otimes e^{\tfrac{z}{2} X_3}}^{(N-2)}
+ \dots \cr
&&\qquad\qquad\qquad\quad\quad + \overbrace{e^{-\tfrac{z}{2} X_3}\otimes
e^{-\tfrac{z}{2} X_3}\otimes\dots\otimes e^{-\tfrac{z}{2} X_3}}^{(N-1)}
\otimes X_\pm.
\label{li}
\eea
Equivalently, this expression can be written from (\ref{buena}) as
\be 
\Dco^{(N)}(X_\pm)=\Dco^{(m)}(X_\pm) \otimes
e^{\tfrac{z}{2}\Dco^{(N-m)}(X_3)} + 
e^{-\tfrac{z}{2}\Dco^{(m)}(X_3)}\otimes \Dco^{(N-m)}(X_\pm),
\label{lm}
\ee
where $m=1, \dots,(N-1)$.
By writing such $N$-th deformed coproducts as
$ 
X_i^{(N)}:=\Delta^{(N)}(X_i),
$ 
the $q$-deformed CG system is defined by the Hamiltonian
\be
H_z^{(N)}:=C_z^{(N)}= \frac{1}{4}\left\{\frac{\sinh(\frac{z}{2}
X_3^{(N)})}{z/2}\right\}^2 + X_+^{(N)}\, X_-^{(N)}
\label{qcg}
\ee
The integrals of the motion for $H^{(N)}$ are again
$ 
X_i^{(N)}\, (i=3,\pm)
$ 
and the sets $C_z^{(m)}$ and $C_z^{(N-m)}$ with $m=1,\dots (N-1)$.

\subsubsection{Evolution equations for the cluster variables}

Now we can rewrite the
$N$-th coproduct (\ref{lm}) in terms of the cluster variables $X_i^{(m)}$
\bea
&& X_3^{(N)} = X_3^{(m)} + X_3^{(N-m)} \cr
&& X_\pm^{(N)} = e^{-\frac{z}{2} \,X_3^{(m)}} \cdot X_\pm^{(N-m)} + X_\pm^{(m)}
\cdot
e^{\frac{z}{2} \,X_3^{(N-m)}}
\label{costcl}
\eea
However, it turns out to be convenient
to introduce a new basis for the $sl_q(2)$ algebra, namely
\be
S_3=X_3,\qquad\qquad S_\pm=e^{-z\,X_3/2}\,X_\pm.
\label{sbasis} 
\ee
In this new the Poisson brackets read
\be
\pois{S_3}{S_\pm} = \pm \,2\,S_\pm; \quad \pois{S_+}{S_-} =
 \frac{\sinh(z S_3)}{z}e^{-z S_3} + 2\,z\,S_+\,S_-,
\label{slq2}
\ee
the $q$-deformed coproduct is:
\bea
&& \Delta (S_3) = S_3 \otimes id + id \otimes S_3 \cr
&& \Delta (S_\pm) = e^{- z\, S_3} \otimes S_\pm + S_\pm \otimes
id
\label{codefs}
\eea
and the deformed Casimir reads:
\be
C_z =\frac{1}{4}\left\{\frac{\sinh(\frac{z}{2} S_3)}{z/2}\right\}^2 + S_+
S_-\,e^{ z\, S_3}.
\label{casuq2}
\ee 
We remark that in the quantum mechanical case, the basis (\ref{sbasis}) is the
suitable one in order to compute analytically the spectrum of some $sl_q(2)$ operators
(see, for instance, \cite{BCh}-\cite{AW}).

In terms of the $N$-th deformed coproducts
\be
S_i^{(N)}:=\Delta^{(N)}(S_i) 
\ee
the $q$-analogue (\ref{qcg}) of the CG system is defined by the Hamiltonian
\be
H_z^{(N)}:=C_z^{(N)}= \frac{1}{4}\left\{\frac{\sinh(\frac{z}{2}
S_3^{(N)})}{z/2}\right\}^2 + S_+^{(N)}\, S_-^{(N)}\,e^{z\,S_3^{(N)}}
\ee
The integrals of the motion for $H^{(N)}$ are again
$ 
 S_i^{(N)}\,(i=3,\pm) 
$ 
together with $C_z^{(m)}$ and  $C_z^{(N-m)}$ $(m=1,\dots,N-1)$.
The new expressions for the $N$-th coproduct on terms of the
cluster variables $S_i^{(m)}$ are:
\bea
&& S_3^{(N)} = S_3^{(m)} + S_3^{(N-m)} \cr
&& S_\pm^{(N)} = e^{- z\,S_3^{(m)}} \cdot S_\pm^{(N-m)} + S_\pm^{(m)}.
\label{codef}
\eea

The set of $3\,(N-1)$ coupled nonlinear evolution equations
\be
{\dot S}_i^{(m)}=\pois{S_i^{(m)}}{H^{(N)}} \qquad m=1,\dots,(N-1)\qquad i=3,\pm
\label{eve}
\ee
can be splitted into $(N-1)$ copies of the following set of three equations 
\bea
&&\back\back {\dot S}_3^{(m)}=2\,(S_+^{(m)}\delta_- -
S_-^{(m)}\delta_+)\,e^{z\,\delta_3}
\cr 
&&\back\back  {\dot S}_+^{(m)}=2\,z\,\delta_-\,e^{z\,\delta_3}\,S_+^{(m)}(\delta_+ -
S_+^{(m)}) -
\frac{\sinh(z\,\delta_3)}{z}\,S_+^{(m)}+\delta_+\,e^{z\,\delta_3}\,\frac{1-e^{-2\,z\,S_3^{(m)}}}{z}
\label{CGq}\\
&&\back\back
{\dot S}_-^{(m)}=
-2\,z\,\delta_+\,e^{z\,\delta_3}\,S_-^{(m)}(\delta_- -
S_-^{(m)}) +
\frac{\sinh(z\,\delta_3)}{z}\,S_-^{(m)}-\delta_-\,e^{z\,\delta_3}\,
\frac{1-e^{-2\,z\,S_3^{(m)}}}{z}
\nonumber
\eea
where in order to eliminate the $S_i^{(N-m)}$ variables we have used the
$\delta_i\equiv S_i^{(N)}$ constants of the motion given by
\bea
&& \delta_3= S_3^{(m)}+S_3^{(N-m)}  \cr
&& \delta_\pm=   e^{- z\,S_3^{(m)}} \cdot S_\pm^{(N-m)} + S_\pm^{(m)}.
\eea
Note that the limit $z\rightarrow 0$ of the equations (\ref{CGq}) reproduce the CG
dynamics (\ref{CG}) and that we have again a constraint between the cluster
variables coming from the constant of the motion given by the $m$-th deformed Casimir
\be
C_z^{(m)}= \frac{1}{4}\left\{\frac{\sinh(\frac{z}{2}
S_3^{(m)})}{z/2}\right\}^2 + S_+^{(m)}\, S_-^{(m)}\,e^{z\,S_3^{(m)}}.
\nonumber
\ee

Thus, once again the collective variables given by the coproduct give rise to a
``separable" deformed motion for each cluster. We also stress that
such cluster dynamics
is a typical Mean-Field dynamics: the evolution equations
are the same for any $m$, and the effect of the remaining degrees of freedom is
hidden in the constants $\delta_i$ given by the $N$-th coproducts of the
generators. In acertain sense, each $m$-th cluster moves within the mean field generated
by the others. This feature can be interpreted as the $q$-analogue of the fact that, at
a quantum mechanical level, the mean-field approximation for the Gaudin hamiltonian is
exact.

\subsubsection{Explicit solution}

In order to solve the cluster equations (\ref{CGq}), let us
perform the following change of variables:
\bea
&& \tilde S_\pm = S_\pm^{(m)}\delta _\mp \\
&& \tilde S_3 = \exp (-2zS_3^{(m)})
\label{tilde}
\eea
We obtain the new equations:
\bea
&& \dot {\tilde{ S_\pm}} = \mp a (\tilde S_\pm)^2 \pm b \tilde S_\pm \pm c
 (1-\tilde S_3)\\
&& \dot {\tilde{ S_3}} = -2 a  \tilde S_3 (\tilde S_+ - \tilde S_-)
\eea
where we have introduced the following constants:
\bea
&& a = 2z \exp (z\delta _3)\\ 
&& b = 2z \delta _- \delta_+ \exp (z\delta_3) - \frac{\sinh (z\delta _3)}{z}\\
&& c = \frac{\delta _+ \delta _- e^{z\delta_3}}{z}
\eea

Summing and subtracting the above equations we get, respectively:
\bea
&& \dot {\tilde{ S_+}} +\dot {\tilde{ S_-}} = - a( \tilde S_+ - \tilde S_-)\left(
 \tilde S_+ +  \tilde S_- -\frac{b}{a}\right) \\
&& \dot {\tilde{ S_+}} -\dot {\tilde{ S_-}} = -a(\tilde S_+^2 + \tilde S_-^2) 
+b(\tilde S_+ + \tilde S_- ) +2c (1- \tilde S_3)
\eea
Thus we have:
\be
\frac{\dot {\tilde{ S_3}}}{\tilde S_3} = -2 a ( \tilde S_+ - \tilde S_-) =
2 \frac{\dot {\tilde{ S_+}}+ \dot {\tilde{ S_-}}}
{\tilde S_+ + \tilde S_-  -\frac{b}{a}}
\ee
which implies 
\be
\tilde S_3 = k_m \left(\tilde S_+ + \tilde S_-  -\frac{b}{a}\right)^2
\ee
where the number $k_m$ is the integration constant that can be expressed in terms of
the values of the
$m$-th cluster Casimirs as follows:
\be
{k_m}^{-1}=\left(C_z^{(m)} -
e^{-z\,\delta_3}\,C_z^{(N-m)}+\frac{1-e^{-z\,\delta_3}}{2\,z^2}    \right)^2 .
\ee

Now it is natural to introduce:
\be
Y_+ = \tilde S_+ + \tilde S_-  -\frac{b}{a}\qquad\quad
Y_- = \tilde S_+ -\tilde S_-
\ee
entailing:
\bea
&& \dot Y_+ = -a\, Y_- Y_+ \\
&& \dot Y_- = -\frac{a}{2}[(1+4\frac{c\,k_m}{a})Y_+^2 + Y_-^2] + \frac{b^2+4ac}{2a}
\eea
Setting:
\bea
&& Z_+ = \frac{a}{(b^2 +4ac)}(1+4\frac{c\,k_m}{a})^{1/2}Y_+\\
&& Z_- =\frac{a}{(b^2 +4ac)} Y_-
\eea
we obtain the evolution equations 
\bea
&& \dot Z_+ = -\Omega\,Z_+\,Z_-\\
&& \dot Z_- = -\frac{1}{2}\,\Omega\,\left(Z_+^2 + Z_-^2 -\frac{1}{\Omega}\right).
\eea
By putting $\Omega=(b^2 +4ac)$, the following equations are obtained
\be
\dot{Z}_+ \pm  \dot{Z}_- =\mp\, \frac{\Omega}{2}\left((Z_+ \pm 
Z_-)^2-\frac{1}{\Omega}\right)
\ee
whose solution  reads:
\bea
Z_+ \pm  Z_- = \pm\frac{1}{\sqrt{\Omega}}\tanh (\frac{\sqrt{\Omega}}{2}(t-t_\pm^{(m)}))
\eea
Note that 
\be
\frac{\sqrt{\Omega}}{2}=\sqrt{C_z^{(N)}+z^2(C_z^{(N)})^2}
\ee
and the limit $z\to 0$ of this expression gives the undeformed period $\sqrt{C^{(N)}}$.
 Two remarks are here in order: first, on a given energy surface belonging  to  the open subset of the phase space $0 > C_z^{(N)} > -z^{-2}$ all the orbits are periodic with the same period; second, the nature of the deformed dynamics turn out to depend explicitly on the value of $z$.

\subsection{A Gaudin system on a $q$-Poincar\'e algebra}

We now consider the Poisson analogue of the
(non-coboundary) quantum deformation of the (1+1) Poincar\'e algebra ${\cal P}(1,1)$ in
terms of ``light-cone" coordinates:
\be
\pois{X_3}{X_\pm} = \pm\,2\, X_\pm  \qquad \pois{X_+}{X_-} = 0
\label{aap}
\ee 
The
deformed coproduct is (see \cite{Rag} with $X_3=2\,K$ and $X_\pm=H\pm P$)
\bea
&& \Delta (X_3) = X_3 \otimes id + id \otimes X_3 \cr
&& \Delta (X_\pm) = e^{-\frac{z}{2}  X_3} \otimes X_\pm + X_\pm \otimes
e^{\frac{z}{2} X_3}
\label{codefpo}
\eea
that, in spite of the non-triviality of the deformation, is still
compatible with the undeformed brackets (\ref{aap}). This deformation is isomorphic to
the one firstly introduced in $\cite{VK}$.
The known Casimir function for ${\cal P}(1,1)$ (and, consequently, for this
deformation) is: 
\be
C_z = X_+ X_-
\ee 

Again, we shall make use of the new basis
\be
S_3=X_3,\qquad\qquad S_\pm=e^{-z\,X_3/2}\,X_\pm. 
\ee
for which  the ${\cal P}(1,1)$ Poisson brackets read
\be
\pois{S_3}{S_\pm} = \pm \,2\,S_\pm \qquad \pois{S_+}{S_-} =
 2\,z\,S_+\,S_-
\label{popoi}
\ee
and the $q$-deformed coproduct is (\ref{codefs}).
The deformed Casimir function is:
\be
C_z = S_+
S_-\,e^{ z\, S_3}.
\label{caspoi}
\ee

We will consider again the ``$q$-Gaudin" Hamiltonian given by the $N$-th coproduct of
the deformed Casimir
\be
H^{(N)}= S_+^{(N)} \,S_-^{(N)}\,e^{z\,S_3^{(N)}}.
\ee
The associated dynamics is given by (\ref{eve}), and 
by using the constants of the motion provided by the $N$-th coproducts 
\bea
&& \delta_3= S_3^{(m)}+S_3^{(N-m)}  \cr
&& \delta_\pm=   \exp(- z\,S_3^{(m)}) \cdot S_\pm^{(N-m)} + S_\pm^{(m)}
\eea
the evolution equations are splitted into $(N-1)$ copies
($m=1,\dots,(N-1)$) of the set:
\bea
&&\back\back {\dot S}_3^{(m)}=2\,(S_+^{(m)}\delta_- -
S_-^{(m)}\delta_+)\,e^{z\,\delta_3}
\cr 
&&\back\back  {\dot S}_+^{(m)}=2\,z\,\delta_-\,e^{z\,\delta_3}\,S_+^{(m)}(\delta_+ -
S_+^{(m)}) \label{PGq}\\
&&\back\back
{\dot S}_-^{(m)}=
-2\,z\,\delta_+\,e^{z\,\delta_3}\,S_-^{(m)}(\delta_- -
S_-^{(m)}) \nonumber
\eea
In the limit $z\rightarrow 0$ these equations reproduce 
\bea
&& {\dot X}_3^{(m)}=2\,(X_+^{(m)}\,\delta_-- X_-^{(m)}\,\delta_+)\cr
&& {\dot X}_+^{(m)}={\dot X}_-^{(m)}=0.
\eea
which leads to the (trivial)
dynamics for the non-deformed Poincar\'e-Gaudin system.

\subsubsection{Solutions for the cluster equations}

The system (\ref{PGq}) can be solved on the very same footing as the qCG system. The
variables (\ref{tilde}) obey now the simpler equations:
\bea
&& \dot {\tilde{ S_\pm}} = \mp a (\tilde S_\pm)^2 \pm b \tilde S_\pm \\
&& \dot {\tilde{ S_3}} = -2 a  \tilde S_3 (\tilde S_+ - \tilde S_-)
\eea
where
\bea
&& a = 2z \exp (z\delta _3)\\
&& b = 2z \delta _- \delta_+ \exp (z\delta _3) =  \delta _- \delta_+ a~ =~
2\,z\,C^{(N)}_z
\eea
The evolution equation for  ${\tilde{ S_\pm}}$ can be immediately integrated, yielding the ``kink-shape'' solution:
\be
 {\tilde{ S_\pm}}= \frac{b}{2a}[1 + \tanh (\mp \frac{b}{2}t-\phi_\pm^{(m)}))]
\label{solgp}
\ee
Consequently, $S_3$ has the following time-behaviour:
\be
S_3^{(m)} = -(z^{-1} \log [(\cosh (bt + \phi _+ -\phi _-) + 
\cosh ( \phi _+ +\phi _-) )/2 - \cosh (\phi_+) \cosh (\phi_-) ] 
\ee
which reduces to a linear function of $t$ in the limit $z \to 0$.
 Note that one could add again a constant (depending on $m$) to the solution; such
constant has to vanish in the limit $z \to 0$.
Recall that the constants of the motion given by the cluster Casimirs 
\be
C_z^{(m)}= S_+^{(m)}\, S_-^{(m)}\,e^{z\,S_3^{(m)}} \qquad \mbox{and} \qquad
C_z^{(N-m)}= S_+^{(N-m)}\, S_-^{(N-m)}\,e^{z\,S_3^{(N-m)}} 
\ee
will be also related to the remaining integration constants.

\subsection{A $q$-Ruijsenaars-Schneider system}

We consider again the $P_z(1+1)$ algebra and the following Hamiltonian 
which under a specific symplectic realization \cite{Rag} resembles the
Ruijsenaars-Schneider system \cite{RS}
\be
H^{(N)}= X_+^{(N)} + X_-^{(N)}
= (S_+^{(N)} + S_-^{(N)})\,e^{z\,S_3^{(N)}/2}.
\label{qRS}
\ee
This Hamiltonian Poisson-commutes with the $N$ Casimirs
\be
C^{(m)}= S_+^{(m)}\, S_-^{(m)}\,e^{z\,S_3^{(m)}}, \qquad
m=1,\dots,N.
\ee
On the other hand, since $\pois{X_+^{(N)}}{X_-^{(N)}}=0$, we have two additional
constants of the motion (the $N$-th Casimir is just the product of them) given by
\be
\delta_\pm= X_\pm^{(N)}=e^{z\,S_3^{(N)}/2}\,S_\pm^{(N)} =e^{-
z\,(S_3^{(m)}-S_3^{(N-m)})/2}
\cdot S_\pm^{(N-m)} + S_\pm^{(m)}\cdot e^{ z\,(S_3^{(m)}+S_3^{(N-m)})/2}.
\ee
Moreover, note that $\delta_+ +\delta_-=H^{(N)}$. 
By using such constants we get the following sets of separated equations for the
cluster variables
$S_i^{(m)}$:
\bea
&& {\dot S}_3^{(m)}=2\,(S_+^{(m)}-
S_-^{(m)})\,e^{z\,S_3^{(N)}/2}
\cr 
&&  {\dot S}_+^{(m)}=z\,S_+^{(m)}\left( (\delta_+ + \delta_-) - 2\, S_+^{(m)}
\,e^{z\,S_3^{(N)}/2} 
\right)\label{qRS}\\ 
&&
{\dot S}_-^{(m)}=-z\,S_-^{(m)}\left( (\delta_+ + \delta_-) - 2\, S_-^{(m)}
\,e^{z\,S_3^{(N)}/2}\right)  \nonumber
\eea
Since it is immediate to check that
\be
{\dot S}_3^{(N)}=2\,(\delta_+ - \delta_-) \quad \longrightarrow
\quad  S_3^{(N)}=2\,(\delta_+ -
\delta_-)t + \beta
\label{lins3}
\ee
through the change of variables
\be
Y_\pm^{(m)}=S_\pm^{(m)}\,e^{z\{(\delta_+ - \delta_-)\,t+\beta\}}
\label{change}
\ee
we get the explicit solution for the variables $Y_\pm^{(m)}$:
\be
Y_\pm^{(m)}=\frac{\delta_\pm}{2}\left\{1 + \tanh \left(\pm
z\,\delta_\pm(t-t_\pm^m)\right)\right\}
\ee
which is again ``kink-shape". Note that, when compared with (\ref{solgp}), the
exponential term (\ref{change}) for the
$S_\pm^{(m)}$ variables and the linear motion for $S_3^{(N)}$ (\ref{lins3}) are the
only dynamical differences between both systems, both of them due to the fact that the
Hamiltonian (\ref{qRS}) does not commute with
$S_3^{(N)}$.

\section{Poisson-Lie dynamics on a solvable group}

The definition of all the cluster variables for the previous systems is always
based on the same type of deformed coproduct (namely (\ref{codef}) and its $S$-basis
form (\ref{codefs})), although each particular dynamics is specialized by a given
Poisson structure and Hamiltonian. As we shall see in the sequel, this observation can
be rephrased in group theoretical terms by considering such coproduct as the group law
for a given solvable Lie group.
Namely, let us consider the following element $g_z$ of a three-dimensional Lie group
$G_z$:
\be
g_z=  \left( 
\begin{array}{ccc}
 e^{-z\,A} & 0 & C\cr
 0 & e^{-z\,A} & B\cr
 0 & 0 & 1
\end{array}
\right) 
\label{int}
\ee 
The group parameters are $A,B$ and $C$ ($z$ is a real constant). It is
straightforward to compute the Lie algebra generators associated to $A,B$ and $C$ 
which are, respectively,
\be
D=  \left( 
\begin{array}{cccc}
 -z & 0 & 0\cr
 0 & -z & 0\cr
 0 & 0 &  0
\end{array}
\right) 
\quad
P_1=  \left( 
\begin{array}{cccc}
 0 & 0 & 0\cr
 0 & 0 & 1\cr
 0 & 0 & 0
\end{array}
\right) 
\quad
P_2=  \left( 
\begin{array}{cccc}
 0 & 0 & 1\cr
 0 & 0 & 0\cr
 0 & 0 & 0
\end{array}
\right) 
\ee
These generators close the following three dimensional solvable Lie algebra
\be
\conm{D}{P_1}=-z\,P_1
\qquad
\conm{D}{P_2}=-z\,P_2
\qquad
\conm{P_1}{P_2}=0
\ee
that contains a dilation $D$ together with two euclidean translations $P_1$ and $P_2$. Obviously, when
$z\neq0$ such parameter can be reabsorbed through the automorphism 
$D\rightarrow D/z$. On the other hand, the $z\rightarrow 0$ limit of $G_z$ is the three
dimensional abelian group (see \cite{BHOSpl} and \cite{L,LM} for the general
connection between quantum algebras and dual Poisson-Lie groups).

A natural coalgebra structure on $\mbox{Fun}(G_z)$ is defined through the coproduct
given by the matrix multiplication on a representation on the group
\be
\Delta(g_z):=g_z\, \dot\otimes \,g_z.
\ee
From the representation (\ref{int}) such a coproduct takes the following values on the
group entries:
\bea
&&\Delta(1)=1\otimes 1\cr
&&\Delta(e^{-z\,A})=e^{-z\,A}\otimes e^{-z\,A}\cr
&&\Delta(B)=e^{-z\,A}\otimes B + B\otimes 1\label{coabc}\\
&&\Delta(C)=e^{-z\,A}\otimes C + C\otimes 1.
\nonumber
\eea
Note that the second expression is equivalent to the fact that 
\be
\Delta(A)=1\otimes A + A\otimes 1.
\ee
Therefore, under the identification
\be
A\equiv S_3 \qquad B\equiv S_+ \qquad C\equiv S_- 
\label{iden}
\ee
the coproduct (\ref{codefs}) is just (\ref{coabc}). Therefore, all the
$N$-th coalgebra systems described above are defined on
$N$ tensor copies of different Poisson algebras of (smooth) functions on $G_z$. In
general, any Poisson structure on
$\mbox{Fun}(G_z)$ for which the group multiplication (\ref{coabc}) is a Poisson map
endows $\mbox{Fun}(G_z)$ with the structure of a Poisson-Lie group
(see, for instance, \cite{Dr}-\cite{PLr} and references therein). Hence, all the
previous results can be interpreted as examples of Poisson-Lie (PL) dynamics on
$\mbox{Fun}(G_z)$. 

\subsection{The PL $sl_q(2)$ dynamics}

Through (\ref{iden}), the $sl_q(2)$ Poisson coalgebra (\ref{slq2}) can be identified as
a Poisson-Lie structure on
$\mbox{Fun}(G_z)$ with the following Poisson brackets
\bea
&& \pois{A}{B}= 2\,B    \cr
&& \pois{A}{C}= - 2\,C    \label{p0}\\
&& \pois{B}{C}= \frac{1-e^{-2\,z\,A}}{z}   + 2\,z\,B\,C ,
\nonumber
\eea
coproduct (\ref{coabc}) and Casimir function (\ref{casuq2})
\be
I_z=\frac{1}{4}\left(\frac{\sinh{z\,A/2}}{z/2}\right)^2 + B\,C\,e^{z\,A}.
\ee
From this perspective, the $q$-CG system is just obtained when we consider the evolution
under 
the particular Hamiltonian
$H^{(N)}=\Delta^{(N)}(I_z)$. Note that the non-deformed CG system is obtained in
thelimit $z\to 0$, that corresponds to the three dimensional abelian group $G_0$ where
the additive group law for the parameters is just the primitive coproduct (\ref{prim}).

\subsection{The PL $q$-Euclidean dynamics}

In the same way, the Poisson $P_z(1+1)$ algebra (\ref{popoi}) can be thought as a
different Poisson-Lie structure on
$\mbox{Fun}(G_z)$ endowed with the Poisson brackets:
\bea
&& \pois{A}{B}= 2\,B    \cr
&& \pois{A}{C}= -2\,C    \label{p1}\\
&& \pois{B}{C}= 2\,z\,B\,C.
\nonumber
\eea
The Casimir function is the analogue of (\ref{caspoi}):
\be
I_z=B\,C\,e^{z\,A}.
\ee
Now, the $q$-Gaudin system on the Poincar\'e algebra is defined by the Hamiltonian
\be
H^{(N)}=\Delta^{(N)}(B\,C\,e^{z\,A})
\ee
whilst the Hamiltonian for the $q$-Ruijsenaars-Schneider system is 
\be
H^{(N)}=\Delta^{(N)}\left((B+C)e^{z\,A/2}\right).
\ee
We stress that both systems are defined with respect to the same PL bracket (\ref{p1}).

\subsection{PL $sl_q^\kappa (2)$ dynamics and contractions}

>From this PL point of view it would be natural to explore the dynamics induced
from other Poisson-Lie structures on $G_z$ and to exploit the PL structure in order to
extract more dynamical information. For instance, we could realize that (\ref{p0}) and
(\ref{p1}) can be simultaneously written as a one-parameter family of PL brackets:
\bea
&& \pois{A}{B}= 2\,B    \cr
&& \pois{A}{C}= - 2\,C    \label{p2}\\
&& \pois{B}{C}= \kappa\,\frac{1-e^{-2\,z\,A}}{z}   + 2\,z\,B\,C 
\nonumber
\eea
where the paramater $\kappa\geq 0$ can be interpreted as a contraction parameter: if
$\kappa=1$ we have the $sl_q(2)$ structure and the limit $\kappa\to 0$ leads to the
$q$-Poincar\'e one. Note that (\ref{p2}) is compatible with the group coproduct
(\ref{coabc}) for any value of $\kappa$.
The Casimir function for this more general bracket is
\be
I_z^\kappa=\frac{\kappa}{4}\left(\frac{\sinh{z\,A/2}}{z/2}\right)^2 + B\,C\,e^{z\,A}.
\ee
It is
easy to check that the equations for the cluster variables (\ref{PGq}) of the
$q$-Gaudin-Poincar\'e system are just the $\kappa\to 0$ contraction of the $q$-CG
equations (\ref{CGq}), provided the latter system is defined by using the Poisson
algebra $sl_q^\kappa (2)$ (\ref{p2}) instead of the $sl_q(2)$ one (\ref{p0}).

Finally we remark that, since the Poisson structure (\ref{p2}) is
quadratic in terms of the entries of $g_z$ (note that the number $1$ is one of
such entries) it is possible to rewrite (\ref{p2}) in the form
\be
\poiss{G_z}{\ G_z}=\conm{r}{G_z\dot\otimes G_z}
\ee
where $r$ is a $9\times 9$ constant classical $r$-matrix. We also recall that, in
general, a suitable quantization of Poisson-Lie groups gives rise to quantum groups
and some quantum versions of different PL brackets on the group
$G_z$ were already given in \cite{BHOSpl}.

\bigskip
\bigskip

\noindent
{\Large{{\bf Acknowledgments}}}

\bigskip

\noindent
 A.B. has been partially supported by MCyT and Junta de Castilla y Le\'on (Espa\~na,
Projects BFM2000-1055 and BU04/03). O.R. has been partially supported by INFN and by MIUR
(COFIN2001 ``Geometry and Integrability").

%\small

\end{document}